\documentclass[pre,aps,twocolumn,superscriptaddress]{revtex4}

\usepackage[dvips]{graphicx}
\usepackage{amssymb,amsfonts,amsmath}
\usepackage{color}
\usepackage{ulem}

\newcommand{\rv}{{\mathbf r}}

\newcommand{\e}{{\rm e}}
\newcommand{\J}{{\bf J}}

\newcommand{\Xv}{{\bf X}}

\newcommand{\vel}{{\bf v}}

\newcommand{\rnt}{{\bf r}^N\!\!,t}
\newcommand{\kT}{k_{\rm B}T}

\newcommand{\GvH}{G_{\rm vH}}
\newcommand{\JvH}{{\bf J}_{\rm vH}}
\newcommand{\JvHf}{{\bf J}^{\rm f}_{\rm vH}}
\newcommand{\JvHb}{{\bf J}^{\rm b}_{\rm vH}}
\newcommand{\JvHfddft}{{\bf J}^{\rm f,ad}_{\rm vH}}
\newcommand{\Jtwo}{{\sf J}_2}

\newcommand{\rns}{{\bf r}^N\!\!,s}
\newcommand{\Omfree}{\hat{\Omega}_{\rm src}}

\begin{document}

\title{Dynamic correlations in Brownian many-body systems}

\author{Joseph M. Brader$^*$}
\affiliation{Soft Matter Theory, 
  University of Fribourg, CH-1700 Fribourg, Switzerland}
\email{Joseph.Brader@unifr.ch}

\author{Matthias Schmidt}
\affiliation{Theoretische Physik II, Physikalisches Institut, 
  Universit{\"a}t Bayreuth, D-95440 Bayreuth, Germany}
\email{Matthias.Schmidt@uni-bayreuth.de}

\date{19 October 2013, accepted: 16 December 2013,
J. Chem. Phys. {\bf 140}, 034104 (2014).}

\begin{abstract}
For classical Brownian systems driven out of equilibrium we derive
inhomogeneous two-time correlation functions from functional
differentiation of the one-body density and current with respect to
external fields.  In order to allow for appropriate freedom upon
building the derivatives, we formally supplement the Smoluchowski
dynamics by a source term, which vanishes at the physical
solution. These techniques are applied to obtain a complete set of
dynamic Ornstein-Zernike equations, which serve for the development of
approximation schemes. The rules of functional calculus lead naturally
to non-Markovian equations of motion for the two-time
correlators. Memory functions are identified as functional derivatives
of a unique space- and time-nonlocal dissipation power functional.
\end{abstract}


\maketitle

\section{Introduction}

In a standard procedure of equilibrium Statistical Mechanics one
generates static correlation functions of interacting many-body
systems by functional differentiation.  For example, differentiating
the position-dependent average one-body density distribution,
$\rho(\rv)$, with respect to an external potential field, $V_{\rm
  ext}(\rv)$, yields the autocorrelation function of density
fluctuations \cite{evans79}
\begin{align}
-\frac{\delta \rho(\rv_1)}{\delta \beta V_{\rm ext}(\rv_2)}
= \langle \hat{\rho}(\rv_1)\hat{\rho}(\rv_2) \rangle - \rho(\rv_1)\rho(\rv_2), 
\label{example}
\end{align} 
where $\hat{\rho}(\rv)=\sum_{i=1}^{N}\delta(\rv-\rv_i)$ is the
classical density operator, $\delta(\cdot)$ is the Dirac distribution,
$\beta=1/(k_{\rm B}T)$, where $k_{\rm B}$ is the Boltzmann constant, $T$ is
absolute temperature, and the angular brackets denote an average over
an appropriate statistical ensemble.  Proving the equality
\eqref{example} is straightforward, but requires as input the explicit
form of the equilibrium (Boltzmann-Gibbs) probability distribution
function in the grand canonical ensemble.

The situation for calculating {\it dynamic} two-body correlation
functions is quite different; there is presently no standard method
for identifying, in the spirit of \eqref{example}, time-dependent
microscopic correlation functions, such as the van Hove function
\cite{vanHove54,Hansen06}, with functional derivatives of average
quantities. In contrast to the static case, calculating dynamic
correlations not only requires knowledge of the many-body probability
distribution at a given time, but requires also the transition
(conditional) probability between two states of the system at
different times.  The transition probability encodes the specific
microscopic dynamics under consideration and is closely related to the
propagator, which generates the time evolution of the distribution
function.  In a very recent publication \cite{brader13noz} we have
exploited this connection in order to obtain two-time correlations
from functional derivatives of the one-body fields, focusing on the
special case of many-body Brownian dynamics and constructing a special
external field for building the derivative.

A general rigorous identification of microscopic time-correlation
functions with corresponding functional derivatives would have
far-reaching consequences.  In particular, such an identification
would be essential when seeking to unify theories formulated on the
one-body level, such as classical dynamical density functional theory
(DDFT) \cite{evans79}, with the numerous approaches that aim to treat
the dynamics of the two-time dynamical correlation functions, such as
mode-coupling theory (MCT)~\cite{goetze_book}. In treating a dynamical
problem one often has access to an equation of motion for the time
evolution of a given one-body average, such as the density or the
current. While in very rare cases this may be exact, one is usually
faced with an approximate expression. However, in both cases,
differentiation of the one-body expression with respect to an
external, time-dependent field constitutes a method to generate an
equation of motion for the inhomogeneous two-time correlations. Here
it is crucial that the functional derivatives involved have been
identified with well-defined microscopic correlation functions.
Clearly, if the one-body `parent' equation of motion is approximate in
nature, then the derived `descendant' equation provides the dynamics
of the two-time correlations on a similar level of description. If the
parent equation is exact, then so too is the equation for the two-body
correlations.  The approach is general; higher-order members of a
complete hierarchy are obtained by further differentiation.

The above strategy has been successfully applied to obtain a
nonequilibrium Ornstein-Zernike (NOZ) equation for overdamped Brownian
systems \cite{brader13noz}.  The static Ornstein-Zernike (OZ) relation
\cite{Hansen06}, utterly familiar from equilibrium liquid-state
theory, is hence generalized to arbitrary dynamical situations,
including bulk dynamics in equilibrium as a non-trivial special
case. The NOZ equation is nonlocal in spacetime and incorporates
memory functions, which play a role for the dynamics analogous to that
of the equilibrium direct correlation function for the static
structure. An unexpected and remarkable result is that the
non-Markovian structure of the mode-coupling equation of motion for
the bulk intermediate scattering function \cite{goetze_book} is
recovered in a natural way, purely as a consequence of applying the
rules of functional calculus.  Furthermore, the present approach sheds
light on the test-particle approach of Archer {\it et.~al}
\cite{dtplhopkins10dtpl}.  It thus appears that the method of
dynamical functional differentiation provides a powerful tool for
generating new dynamical theories self-consistent on the one- and
two-body level, as well as for extending existing theories of the
one-body functions (such as DDFT) to the two-body level.

The restricted set of two-body correlation functions described in
Ref.~\cite{brader13noz} were obtained by functional differentation of
the (Smoluchowski) propagator with respect to the special choice of
one-body field
\begin{align}
 \mathcal{V}(\rv,t) &\equiv \int_{t_0}^t dt' D_0\nabla^2 V_{\rm
    ext}(\rv,t'),
 \label{curlyV}
\end{align}
where $V_{\rm ext}(\rv,t)$ is the (time-dependent) external potential,
$D_0$ is the bare diffusion coefficient and $t_0$ is an initial
time. Equation \eqref{curlyV} amounts to the application of the
diffusion operator to the external potential, which is necessary in
order to ``co-evolve'' the external potential along the physical
dynamics of the system.  The choice to employ the field \eqref{curlyV}
was motivated by the desire to obtain the most direct mathematical
route to the NOZ equation, but sacrificed the full generality of the
approach.

In the present paper we address the general case by differentiating
with respect to the bare external fields, i.e.\ a non-conservative
force field $\Xv(\rv,t)$ and the temporal rate of change of an
external potential, $\dot V_{\rm ext}(\rv,t)$ (which is in
time-dependent situations a more natural quantity than the external
potential itself).  A key concept required for these calculations is
that of a `sourced dynamics', which can formally differ from the
(Smoluchowski) dynamics of the physical system. The sourced dynamics
is defined by a non-standard, many-body time evolution operator on
configuration space which is not constrained by the many-body
continuity equation. Physically meaningful two-point correlation
functions are generated from functional differentiation of the sourced
dynamics propagator, and imposing the continuity equation {\it a
  posteriori}. Following this procedure, we generate a complete set of
dynamic correlation functions, the most important of which is the
tensorial two-body current, auto-correlating the microscopic particle
current at two different points in spacetime.  Using the two-body
current we derive the most general NOZ equation for Brownian dynamics.

The paper is organized as follows: In Sections
\ref{OneBodyAverages}-\ref{MicroscopicDynamics} we define the relevant
correlation functions and specify the microscopic dynamics of
interest.  In Sec.~\ref{TwoTimeAverages} we give a detailed
description of how two-time correlation functions are obtained as
averages over configuration space. The sourced dynamics introduced in
Sec.~\ref{FreeDynamics} is used in Sec.~\ref{DynamicalDerivatives} to
obtain two-time correlations, which are then
(Sec.~\ref{ResponseFunctions}) interpreted physically as response
functions.  We next apply our strategy to develop NOZ equations for
the two-time correlations, applying first an adiabatic approximation
(Sec.~\ref{DDFTapproximation}), before proceeding to develop an exact
superadiabatic expression involving `time-direct correlation
functions', which are memory functions that account for structural
relaxation (Sec.~\ref{beyondDDFT}).  In Sec.~\ref{TheConnection} we
show that within the recently developed power functional theory
\cite{power}, the time-direct correlation functions can be identified
as functional derivatives of an excess (over ideal gas) power
dissipation functional. A significant implication is that by
approximating a single generating functional one can construct a
consistent, non-adiabatic theory for both the one- and two-body
nonequilibrium correlations.  In Section \ref{conclusions} we give
concluding remarks and an outlook on future work.

\section{Theory}

\subsection{One-body correlations}\label{OneBodyAverages}
For a classical many-body system subject to arbitrary microscopic
dynamics, the one-body density and one-body current are described by
the operators
\begin{align}\label{density_operator}
  \hat\rho(\rv,t) &= \sum_i \delta(\rv-\rv_i),\\
  \label{current_operator}
  \hat\J(\rv,t) &= \sum_i \delta(\rv-\rv_i) \hat\vel_i(t),
\end{align}
respectively, where $\hat\vel_i(t)$ is the time-dependent velocity of
particle $i$ and the sum runs over all particles, $i=1,\ldots,N$.  The
mathematical character of the velocity appearing in
\eqref{current_operator} depends upon whether one adopts a trajectory
based (Newtonian or Langevin) picture of the dynamics, or a
probabilistic phase space (Liouville or Smoluchowski) picture.  In the
case of the probabilistic interpretation of overdamped Langevin
dynamics, which we will adopt henceforth, the particle velocity is a
differential operator on configuration space.

The local conservation of particle number is expressed by the one-body
continuity equation
\begin{align}
  \frac{\partial}{\partial t_1} \rho(1)
  &= -\nabla_1\cdot \J(1), 
  \label{continuity1}
\end{align}
where the average density and current are given by
$\rho(1)=\langle\hat\rho(1)\rangle$ and
$\J(1)=\langle\hat\J(1)\rangle$, respectively. Here the angular
brackets indicate a statistical average with respect to the
appropriate distribution function, as specified below.  We have
introduced for brevity the shorthand notation $\hat\rho(1) \equiv
\hat\rho(\rv_1,t_1)$, and $\hat\J(1) \equiv \hat\J(\rv_1,t_1)$ for
spacetime points, and $\nabla_1$ indicates the derivative with respect
to~$\rv_1$. The average one-body velocity is simply
$\vel(1)=\J(1)/\rho(1)$.

\subsection{Two-body correlations}\label{TwoBodyCorrelations}
The most commonly studied dynamical two-body correlation function is
the density-density correlation function introduced by van Hove
\cite{vanHove54,Hansen06}. For spatially and temporally inhomogeneous
situations, the van Hove function is defined as
\begin{align}
  \GvH(1,2) & = \rho(1)^{-1}
  \langle \hat\rho(1)\hat\rho(2) \rangle,
  \label{EQGvH}  
\end{align} 
where the angular brackets indicate an appropriately defined (see
below) two-time average over the nonequilibrium system, which evolves
from the state at the earlier time $t_2$ to the later time $t_1$.

While the van Hove function is very useful for characterizing
relaxation to equilibrium, in general, as a scalar function it
does not provide a complete picture of the particle
dynamics. Additional information is provided by the nonequilibrium
two-body functions
\begin{align}  
  \JvHf(1,2) &= \langle \hat\J(1)\hat\rho(2)\rangle,
  \label{EQJvH}  \\
  \JvHb(1,2) &= \langle \hat \rho(1)\hat\J(2) \rangle,
  \label{EQJvHprime}
\end{align}
which we will henceforth refer to as the front \cite{brader13noz} and
back van Hove current, respectively, and we adopt the causality
convention $t_1\geq t_2$. The two correlation functions are not
equivalent in general, because the particle current is a differential
operator. Note further that the van Hove currents are particularly
important for describing the dynamics in driven systems, such as
e.g.\ in the presence of a time-dependent external potential or
non-conservative shear forces.

The analogue of \eqref{continuity1} on the two-body level is
constituted by two distinct continuity equations
\begin{align}
  \frac{\partial}{\partial t_1} \rho(1) \GvH(1,2)
  &= -\nabla_1\cdot \JvHf(1,2),\label{continuity2}\\
  \frac{\partial}{\partial t_2} \rho(1) \GvH(1,2)
  &= -\nabla_2\cdot \JvHb(1,2), \label{continuity_back}
\end{align}
which relate each vectorial van Hove current to the scalar
van Hove function.

The most fundamental dynamic pair function, however, is the two-body
current
\begin{align}
  \Jtwo(1,2) &= \langle \hat\J(1)\hat\J(2)\rangle. 
  \label{EQJ2definition}
\end{align}
This current-current correlation function is a second-rank tensor
obtained by averaging the dyadic product of two one-body current
operators.  As we shall demonstrate below in
Sec.~\ref{DynamicalDerivatives}, this tensorial correlation function
can be related to a functional derivative of the current with respect
to the nonconservative external force.  Moreover, we will show that
both the front and back van Hove currents, as well as the van Hove
function, can be identified with functional derivatives.

From the two-body current the front and back van Hove currents can be
determined, according to the continuity equations
\begin{align}
  \frac{\partial}{\partial t_1}\JvHb(1,2) &= -\nabla_1\cdot\Jtwo(1,2),
\label{JtwoContinuity1}\\
  \frac{\partial}{\partial t_2}\JvHf(1,2) &= -\nabla_2\cdot\Jtwo(1,2).
\label{JtwoContinuity2}
\end{align}
The two-body current is thus fundamental for studying the dynamics of
liquids, as all pair correlations of lower tensorial rank, namely
\eqref{EQGvH}, \eqref{EQJvH} and \eqref{EQJvHprime}, can be obtained
by building the appropriate divergence in space and integrating in
time.  We demonstrate below (see Secs.~\ref{DDFTapproximation} and
\ref{beyondDDFT}) that focusing on the two-body current enables the
formulation of a general NOZ theory of the dynamic pair correlations.

\subsection{Microscopic dynamics}\label{MicroscopicDynamics}
We next specify the microscopic dynamics with which we will be
concerned for the remainder of the paper. The state of the system is
described by a time-dependent distribution function, $\Psi(\rnt)$,
which gives the probability density to find the $N$ particles in the
system at positions $\rv^N\equiv\{\rv_1,\ldots,\rv_N\}$ at time
$t$. The total interparticle interaction potential is $U(\rv^N)$ and
the particles interact with their surrounding via an external
potential $V_{\rm ext}(\rv,t)$ and via a non-conservative force field
$\Xv(\rv,t)$.  We consider Brownian particles which undergo stochastic
motion and are subject to a velocity-dependent friction force with
force constant~$\gamma$. The overdamped dynamics can be described via
the continuity equation for the many-body distribution function,
\begin{align}
  \frac{\partial}{\partial t}\Psi(\rnt) = 
  -\sum_i \nabla_i\cdot \hat\vel_i(t) \Psi(\rnt).
  \label{smoluchowski}
\end{align}
where the velocity operator of particle $i$ is defined as
\begin{align}\label{vel_op}
  \hat\vel_i(t) = \gamma^{-1}\Big[&-(\nabla_i U(\rv^N))
     - \kT \nabla_i \notag\\&
  -(\nabla_i V_{\rm ext}(\rv_i,t))
  +\Xv(\rv_i,t)  \Big].
\end{align}
Here only the thermal term constitutes a differential operator;
  the two bracketed gradients each yield a vector-valued function,
  which then acts via multiplication only.  Physically, the action of
$\hat\vel_i(t)$ on the distribution, $\Psi(\rnt)$, generates the noise
averaged velocity of particle $i$.

Introducing the Smoluchowski operator \cite{dhont}, which is defined as
\begin{align}\label{EQomegaDefinition}
  \hat{\Omega}(\rv^N\!,t) &= -\sum_i \nabla_i\cdot\hat\vel_i(\rv^N\!,t).
\end{align} 
allows to write the Smoluchowski equation \eqref{smoluchowski} in the
alternative form
\begin{align}
  \frac{\partial}{\partial t}\Psi(\rnt) = 
  \hat\Omega(\rv^N\!,t) \Psi(\rnt).
  \label{smoluchowski_operatorform}
\end{align}
The significant benefit of the rewriting is that a formal solution can
be expressed as
\begin{align}
  \Psi(\rnt) = \e_+^{\int_{t_0}^t \!ds\, \hat{\Omega}(\rv^N\!,s)}\Psi(\rnt_0),
\label{smol_propagator}  
\end{align}
where $\e_+$ indicates the time-ordered exponential operator
\cite{vankampen,braderMCT2012}, which acts on all functions to the
right, and $t_0$ is an initial time, at which the many-body
distribution is assumed to be known. The one-time average of an
operator $\hat f(\rnt)$ on configuration space is thus given by
\begin{align}
f(t) &= \int d\rv^N \hat f(\rnt) \Psi(\rnt),\\
 &=\int d\rv^N \hat f(\rnt)\,\e_+^{\int_{t_0}^t \!ds\,
  \hat{\Omega}(\rv^N\!,s)}\Psi(\rnt_0),
\label{average_expression}
\end{align}
where \eqref{average_expression} follows from substitution of
\eqref{smol_propagator}. Here and in the following, we have suppressed
in the notation the possible dependence of a (dummy) operator $\hat
f(\rnt)$ on further arguments, such as on position $\rv$.

\subsection{Two-time averages}\label{TwoTimeAverages}
We first recall that the correlation between two physical
  quantities, as represented by their corresponding operators $\hat f$
  and $\hat g$ on configuration space, is defined according to
\begin{align}
  C_{fg}(t,t')=\!\int \!d\rv^N\!\!\int 
  \!d\rv'^N\,\hat f(\rnt)\hat g(\rv'^N\!,t')\Psi_2(\rv^N\!,t;\rv'^N\!,t'),
  \label{CfgDefinition}
\end{align}
where $\Psi_2(\rv^N\!,t;\rv'^N\!,t')$ is the joint probability to
find configuration $\rv^N$ at time $t$ and configuration $\rv'^N$ at
time $t'$. Introducing the conditional (transition) probability
$w(\rnt |\,\rv'^N\!,t')$ to find the system in the unprimed state,
given that it was in the primed state at the earlier time~$t'$, the
joint probability can be alternatively expressed as the product
\begin{align}
  \Psi_2(\rv^N\!,t;\rv'^N\!,t')=
  w(\rv^N\!,t\,|\,\rv'^N\!,t')\Psi(\rv'^N\!,t'),
\label{EQpsi2viaTransitionProbability}
\end{align}
where $\Psi(\rv'^N\!,t')$ is, as before, the instantaneous many-body
distribution function. Note that for the case that the system was in
equilibrium at the earlier time~$t'$, such that
$\Psi(\rv'^N,t')=\Psi_{\rm eq}(\rv'^N)$, it is only via the transition
probability that the specific details of microscopic dynamics (whether
e.g.\ Newtonian or Brownian) enter the expression
\eqref{CfgDefinition} for the two-point correlator.

For many-body Brownian dynamics the transition probability evolves in
time according to the (forward) time evolution equation \cite{riskin}
\begin{align}
\frac{\partial}{\partial t} w(\rv^N\!,t\,|\,\rv'^N\!,t') 
= \hat{\Omega}(\rv^N,t)w(\rv^N\!,t\,|\,\rv'^N\!,t'),
\label{evolution1}
\end{align}
which is formally analogous to the equation of motion
\eqref{smoluchowski_operatorform} for the distribution function.  The
formal solution of \eqref{evolution1}, subject to the initial
condition $w(\rv^N\!,t'|\,\rv'^N\!,t')=\delta(\rv^N\!-\rv'^N)$, is
given by
\begin{align}
  w(\rnt\,|\,\rv'^N\!,t') = \e_+^{\int_{t'}^t ds\hat\Omega(\rns)}
  \delta(\rv^N-\rv'^N).
\label{EQpsi2FormalSolution}
\end{align}
Inserting this expression into
\eqref{EQpsi2viaTransitionProbability} allows to rewrite
\eqref{CfgDefinition}, upon carrying out the integral over the primed
coordinates, as
\begin{align}
  C_{fg}(t,t')=\!\int \!d\rv^N\,\!\hat f(\rnt)\,
  \e_+^{\int_{t'}^t \!ds\, \hat{\Omega}(\rv^N\!\!,s)}
  \hat g(\rnt')\Psi(\rv^N\!\!,t').
  \label{EQtwoTimeCorrelation}
\end{align}
The benefits of the propagator form \eqref{EQtwoTimeCorrelation} over
the bare definition \eqref{CfgDefinition} are that only a single
integral over configuration space appears and that only the one-state
distribution $\Psi$ is required, rather than $\Psi_2$. The complexity
of the correlator is hence condensed into the propagator. 
This offers the advantage that all forces appear explicitly,
cf.~the definition \eqref{EQomegaDefinition}.  The structure of
equation \eqref{EQtwoTimeCorrelation} can hence be exploited
\cite{brader13noz} to calculate dynamical functional derivatives. In
the following we seek to develop a general method by which the
inhomogeneous two-time functions \eqref{EQGvH}--\eqref{EQJvHprime} and
\eqref{EQJ2definition} can be connected, by means of time-dependent
functional differentiation (as laid out in
  Sec.~\ref{DynamicalDerivatives}), to the one-body level of
description provided by the density, $\rho(1)$, and current, $\J(1)$.

Our approach to connecting one- and two-time correlators is based on
the fact that taking a functional derivative of
\eqref{average_expression} with respect to an external field appearing
in the propagator (i.e.~the time-ordered exponential) can generate
expressions with the same form as the right hand side of
\eqref{EQtwoTimeCorrelation}.  For particular choices of $\hat f$ and
$\hat g$ and of external field we can thus generate mathematical
relations between physically meaningful one- and two-body correlators.
Remarkably, within this approach the many-body Smoluchowski
propagator, which was solely introduced to evolve the distribution
function in time, cf.~\eqref{smoluchowski_operatorform}, acts as a
dynamical analogue of the Boltzmann factor for static properties in
equilibrium.

Previous work focused on derivatives with respect to the special
choice of external field \eqref{curlyV} and did not reveal the full
generality of dynamical functional differentiation. In the following
we present the complete picture and develop a general method for
obtaining inhomogeneous two-time correlations from the one-body
functions. This however, requires modifying the underlying many-body
dynamics.

\subsection{Generalized many-body sourced dynamics}\label{FreeDynamics}
Here we construct a formal and more general time evolution operator,
which retains the physical Smoluchowski dynamics
\eqref{smol_propagator} as a special case. As a motivation we first
rewrite \eqref{EQomegaDefinition}~as
\begin{align}
\hat\Omega(t)  &= \frac{\beta\gamma}{2}
 \sum_i\left( 
 \hat{\bf v}_i(t)^2
 -\hat{\bf v}_i^{\dagger}(t)\cdot\hat{\bf v}_i(t) 
  \right),
\label{QuantumFieldTheory}
\end{align}
where for arbitrary functions, $a$ and $b$, on configuration space the
adjoint velocity operator $\hat\vel_i^\dag(t)$ obeys
\begin{align}
\int d\rv^N a(\rv^N) \,\hat{\bf v}_i(t)\, b(\rv^N)= \int d\rv^N b(\rv^N) \,\hat{\bf v}_i^\dag(t)\, a(\rv^N). 
\end{align} 
As an aside,  integration by parts yields the identity
\begin{align}
\hat{\bf v}_i^{\dagger}(t)= \hat\vel_i(t)+2D_0\nabla_i, 
\end{align} 
where $D_0=k_{\rm B}T/\gamma$ is the bare diffusion coefficient.
Recognizing the self-adjoint nature of the second contribution in
\eqref{QuantumFieldTheory}, we combine this together with external
contributions into a source operator, defined as
\begin{align}
\hat S(t) \!=\! 
\beta \sum_i \left(\frac{\gamma}{2}\, \hat\vel_i^\dag(t)\!\cdot\!\hat\vel_i(t) 
\!-\! \dot{V}_{\rm ext}(\rv_i,t)\! +\! \alpha(\rv_i,t) \right)
\!+\!\lambda(\rnt),
\label{EQsourceOperator}
\end{align}
where $\dot V_{\rm ext}(\rv,t)$ indicates the (partial) time
derivative of the external potential, $\alpha(\rv,t)$ is a scalar
field of space and time, which represents a local one-body thermostat
that either adds or removes power from the system, and $\lambda(\rnt)$
is an external many-body probability source field that is used below
to restore the local particle conservation.  Adding the source to the
Smoluchowski operator, we arrive at a time evolution operator defined
by
\begin{align}
  \label{EQomegaFree}
  \Omfree(t) &= \hat\Omega(t) + \hat S(t)\\
  &= \beta\sum_i \left(\frac{\gamma}{2}\hat\vel_i(t)^2 
  - \dot V_{\rm ext}(\rv_i,t) + \alpha(\rv_i,t)
  \right)+\lambda(\rnt),\notag
\end{align}
which is free from the constraint of local particle number
conservation. This constraint can easily be reinstated by requiring
the source contribution to vanish, and hence the probability source
field $\lambda(\rv^N,t)$ to satisfy
\begin{align}
  \hat S(t) \psi(\rnt) = 0,
  \label{EQsourceZero}
\end{align}
in which case the Smoluchowski and the sourced evolution operators
become identical, $\hat\Omega=\Omfree$.

On the basis of \eqref{EQomegaFree}, an alternative to
\eqref{smol_propagator} is a propagator that acts according to
\begin{align}\label{free_dynamics}
  \Psi(\rnt)=e_+^{\int_{t_0}^tds \,\Omfree(s)}\Psi(\rnt_0).
\end{align}
Since the Smoluchowski equation \eqref{smoluchowski} is simply the
many-body continuity equation, integration of its formal solution,
i.e.\ of the right hand side of \eqref{smol_propagator}, over all
particle coordinates yields unity for all times.  This expresses the
fact that the many-body distribution function is normalized at all
times.  However, the configurational integral over the right hand side
of \eqref{free_dynamics} is in general, i.e.\ when $\lambda(\rnt)$ is
prescribed such that \eqref{EQsourceZero} is not satisfied, not
constant unity, but rather a function of time.  We define the negative
of this quantity as
\begin{align}\label{partition_function}
  {\mathcal R}(t)=-\!\int d\rv^N  
  e_+^{\int_{t_0}^tds \,\Omfree(s)}\Psi(\rnt_0), 
\end{align}
which carries an implicit functional dependence on the external fields
$\alpha(\rv,t)$ and $\dot{V}_{\rm ext}(\rv,t)$, as well as, via
\eqref{vel_op}, on $V_{\rm ext}(\rv,t)$ and $\Xv(\rv,t)$.  As we show
in the following, the functional \eqref{partition_function} plays the
role of a generator, in the sense that taking functional derivatives
with respect to the external fields, followed by restoration of the
physical Smoluchowski dynamics via \eqref{EQsourceZero}, creates
one-body averages of interest.

In order to illustrate the sourced dynamics further, we write
\eqref{free_dynamics} in differential form
\begin{align}
  \frac{\partial}{\partial t}\Psi(\rnt)
  &=\left( \hat{\Omega}(t) + \hat{S}(t) \right)\Psi(\rnt)
  \label{eom_source}
\end{align} 
where the $\hat S(t)$ is given by \eqref{EQsourceOperator}.  Equation
\eqref{eom_source} makes explicit that, in general, the continuity
equation is violated by the introduction of the additional source term
into the Smoluchowski equation of motion \eqref{smoluchowski}. Only
for the case when the source term vanishes, we recover the physical
dynamics.  When expressed in the form \eqref{eom_source} the modified
dynamics becomes analogous to the Lindblad form of the quantum
mechanical master equation, which describes the non-unitary time
evolution of a reduced density matrix. In a quantum mechanical
context, the appearance of source contributions in the time evolution
equation represents interaction with the unresolved degrees of freedom
which constitute the environment in which the system is embedded.  In
the present context, the use of the source term allows to enforce the
many-body continuity equation, while allowing the necessary freedom
from this constraint upon the ``virtual'' changes that are represented
by the functional derivatives.

\subsection{Dynamic functional derivatives}\label{DynamicalDerivatives}
In order to perform dynamical functional differentiation, we use the
chain rule for time-ordered exponentials, which in the form of a
general operator identity reads
\begin{align}\label{propagator_deriv}
\frac{\delta}{\delta u(\rv,t)}\e_+^{\int_{t_1}^{t_2} \!ds \,\hat{O}(s)}=
\int_{t_1}^{t_2}\!\!ds\,
  \e_+^{\int_{s}^{t_2} \!ds'\, \hat{O}(s')}
  \frac{\delta \hat{O}(s)}{\delta u(\rv,t)}
  \, \e_+^{\int_{t_1}^{s} \!ds'' \hat{O}(s'')},
\end{align}
where $\hat{O}(s)$ is a time-dependent operator with a functional
dependence upon a field $u(\rv,t)$. Using that for time-dependent
fields $\delta u(\rv,t)/\delta u(\rv',t') =
\delta(\rv-\rv')\delta(t-t')$, and that furthermore one can obtain
straightforwardly
\begin{align}
 \frac{\delta\Omfree(s)}{\delta\beta\Xv(\rv,t)} &= \hat\J(\rv,t)\delta(t-s),\\
 \frac{\delta\Omfree(s)}{\delta\beta\alpha(\rv,t)} &= \hat\rho(\rv,t)\delta(t-s),
\end{align}
we thus derive
\begin{align}
  -\frac{\delta {\mathcal R}(t)}{\delta \beta\Xv(\rv,t)} &= \J(\rv,t),
  \label{EQJhatAsFunctionalDerivative}\\
  -\frac{\delta {\mathcal R}(t)}{\delta \beta\alpha(\rv,t)} &= \rho(\rv,t),
  \label{EQrhohatAsFunctionalDerivative}
\end{align}
where we have used the fact that $\Xv(\rv,t)$ is nonconservative,
i.e.\ divergence-free.

Equations \eqref{EQJhatAsFunctionalDerivative} and
\eqref{EQrhohatAsFunctionalDerivative} suggest that well-defined
two-time correlations can be obtained by further functional
differentiation of \eqref{partition_function} with respect to the
conjugate fields. The time-ordered exponential (propagator) in
\eqref{free_dynamics} and the generating functional
\eqref{partition_function} then play a role similar to that of the
Boltzmann factor and partition function, respectively, in equilibrium
Statistical Mechanics.

Using \eqref{propagator_deriv}, it is straightforward to show that the
following functional derivative relations hold
\begin{align}
  &\frac{\delta \J(1)}{\delta \beta\Xv(2)} =
   D_0 \rho(1)\delta(1,2){\bf 1} + \Jtwo(1,2),\label{EQJbyX}\\
  &\frac{\delta \rho(1)}{\delta \beta\Xv(2)} = 
  \JvHb(1,2)\label{EQrhoByX},  
\end{align}
where causality requires $t_2\leq t_1$, our notation is such
that $\delta(1,2)=\delta(\rv_1-\rv_2)\delta(t_1-t_2)$, and
\eqref{EQJbyX} is of second and \eqref{EQrhoByX} of first tensor rank.
Furthermore, analogous reasoning and an additional integration by
parts yields the derivatives with respect to $\alpha(\rv,t)$, or
equivalently with respect to the time derivative of the external
potential,
\begin{align}
  & \frac{\delta\J(1)}{\delta \beta\dot V_{\rm ext}(2)}  =
  D_0\rho(1)\nabla_2\delta(1,2) + \JvHf(1,2),
  \label{EQJbyRho}\\
  & \frac{\delta\rho(1)}{\delta \beta\dot V_{\rm ext}(2)} =
  \rho(1)\GvH(1,2).\label{EQrhoByRho}
\end{align}
We have thus succeeded in identifying the inhomogeneous two-time
correlation functions \eqref{EQGvH}, \eqref{EQJvH},
  \eqref{EQJvHprime}, and \eqref{EQJ2definition} as time-dependent
functional derivatives of the one-body fields.

\subsection{Response functions}\label{ResponseFunctions}
The results for the dynamical functional derivatives, equations
\eqref{EQJbyX}-\eqref{EQrhoByRho}, gain further physical significance
when regarded as dynamic and in general nonlinear response
functions.  Consider first a perturbation of the current,
$\delta\J(1)$, which is generated in response to a change in the
externally applied nonconservative force, $\delta\Xv(2)$.
Mathematically, this may be expressed via the integral relation
\begin{align}
\delta \J(1) = \int d2\, 
\frac{\delta \J(1)}{\delta \Xv(2)}  \cdot\delta\Xv(2), 
\label{response1}
\end{align}
where the integral over spacetime point 2 is to be calculated up to
the present (i.e.\ the one on the left hand side of
  \eqref{response1}) time, according to $\int d2 =\int
d\rv_2\int_{-\infty}^{t_1} dt_2$.  Substitution of \eqref{EQJbyX} into
\eqref{response1} yields
\begin{align}
\delta \J(1) = \frac{\rho(1)}{\gamma}\delta\Xv(1) \;+\;
 \beta\!\int d2\, 
 \Jtwo(1,2)\cdot\delta\Xv(2),
\label{responseJfromX}
\end{align}
which shows that the current perturbation consists of an instantaneous
response to the local force acting at $\rv_1$ (first term) and a
retarded contribution (second term), arising from forces acting at
more distant locations.  The two-body current thus plays the role of a
response function, which is closely related to the `creep compliance'
employed in macroscopic rheology studies to calculate the deformation
of a material caused by an applied stress field \cite{larson}.
Equation \eqref{responseJfromX} is nonlinear and exact, because the
two-body current is itself a functional of the external forces. Linear
response would be recovered by replacing $\Jtwo(1,2)$ with the
translationally invariant equilibrium function, $\Jtwo^{\rm eq}(1-2)$.
In contrast to the instantaneous current response, which is aligned
with the force perturbation, the retarded contribution (the integral
term in \eqref{responseJfromX}) is in general not parallel to
$\delta\Xv(2)$, due to mediated interactions.

A similar approach may be applied to calculating the change in density
$\delta\rho(1)$ arising from a perturbation in the time-dependent
external field, $\delta\dot V_{\rm ext}(2)$. In this case the
appropriate integral relation is given by
\begin{align}
  \delta \rho(1) = \int \!d2 
\frac{\delta \rho(1)}{\delta \dot V_{\rm ext}(2)}
 \delta\dot V_{\rm ext}(2). 
\label{response3}
\end{align}
Insertion of equation \eqref{EQrhoByRho} yields
\begin{align}
\delta \rho(1) = \beta\rho(1)\int d2 
\,\GvH(1,2)\,\delta\dot V_{\rm ext}(2),
\label{responseRhoFromVextdot}
\end{align}
which establishes that the van Hove function converts changes in the
external potential rate into changes in the one-body density
distribution.

Furthermore, expressions involving the van Hove currents are obtained
by considering
\begin{align}
  \delta \rho(1) &= \int d2\, \frac{\delta \rho(1)}{\delta \Xv(2)} 
  \cdot \delta \Xv(2),\\
  \delta \J(1) &= \int d2\, \frac{\delta \J(1)}{\delta \dot V_{\rm ext}(2)}
  \delta \dot V_{\rm ext}(2),
\end{align}
from which, upon considering \eqref{EQrhoByX} and \eqref{EQJbyRho}, one obtains
\begin{align}
  \delta \rho(1) &= \int d2\, \JvHb(1,2)\cdot\delta \Xv(2)
  \label{responseRhoFromX}\\
  \delta \J(1)  &= -\frac{\rho(1)}{\gamma} \nabla_1\dot V_{\rm ext}(1)
  + \beta \int d2\, \JvHf(1,2) \delta\dot V_{\rm ext}(2),
\label{responseJfromVextdot}
\end{align} 
As in the case of \eqref{responseJfromX}, the relationship
\eqref{responseJfromVextdot} consists of a direct and a mediated
contributions, whereas \eqref{responseRhoFromX} has no direct term.

The relations \eqref{responseJfromX}, \eqref{responseRhoFromVextdot},
\eqref{responseRhoFromX}, and \eqref{responseJfromVextdot} provide a
physical interpretation of the two-time correlators as mediators of
changes in both external fields into resulting changes in the one-body
density and current distribution.  A very familiar case is that of
linear response around the equilibrium state, where $\Xv=0$ and $\dot
V_{\rm ext}=0$. In general, however, the perturbations can be around
any trajectory of the system, which is driven out of equilibrium by
external forces.

\subsection{Adiabatic approximation}\label{DDFTapproximation}
The simplest theory for the average one-body current of interacting
particles is the adiabatic dynamical density functional theory (DDFT),
as originally proposed by Evans \cite{evans79}, and subsequently
rederived by both Marconi and Tarazona \cite{marinibettolomarconi99}
and Archer and Evans \cite{archer04ddft}.  DDFT assumes that the
current,
\begin{align}
  \J_{\rm DDFT}(1) = \frac{\rho(1)}{\gamma}\left(
  -\nabla \frac{\delta F[\rho]}{\delta \rho(1)} - \nabla V_{\rm ext}(1) +\Xv(1)
  \right), 
  \label{ddftCurrent}
\end{align}
arises from a balance between forces generated by viscous friction,
$\gamma\vel(1)$, external fields, thermal motion and interparticle
interactions.  The latter two contributions are generated from the
intrinsic Helmholtz free energy functional $F[\rho]$.  Substitution of
\eqref{ddftCurrent} into the one-body continuity equation
\eqref{continuity1} yields a closed drift-diffusion equation for
$\rho(1)$, which is local in time (i.e.\ Markovian
  \cite{zwanzig_book}).

The functional derivative formalism developed above allows to
generate two-point correlation functions within the adiabatic
approximation.  Using \eqref{ddftCurrent} to calculate the functional
derivative $\delta \J(1)/\delta \Xv(3)$, employing the functional
chain rule, and the relations \eqref{EQJbyX} and \eqref{EQrhoByX}
generates an adiabatic approximation to the two-body current,
\begin{align}\label{EQJ2inDDFT}
&\Jtwo^{\rm ad}(1,3)= {\bf v}(1)\JvHb(1,3)
\notag\\
&-D_0\rho(1) \nabla_1 \left(\frac{\JvHb(1,3)}{\rho(1)}
 -\int \!d\rv_2 \,c(1,2_1)\JvHb(2_1,3)\right),
\end{align}
where the equilibrium direct correlation function is the second
functional derivative of the excess (over ideal gas) part of the
intrinsic Helmholtz free energy, $c(\rv_1,\rv_2)\!=\!- \delta^2 \beta
F^{\rm exc}[\,\rho]/\delta\rho(\rv_1)\delta\rho(\rv_2)$
\cite{evans79}.  The argument $2_1$ in \eqref{EQJ2inDDFT} indicates
position $\rv_2$ and time $t_1$; the direct correlation function is
hence evaluated at distinct values of the spatial arguments, but at
the same time, i.e.\ $c(1,2_1)\equiv c(\rv_1,\rv_2,t_1)$, and ${\bf
  v}(1)$ is given here by $\J_{\rm DDFT}(1)/\rho(1)$.  The three
contributions to \eqref{EQJ2inDDFT} represent a transport term, ideal
decay, and an adiabatic spatial convolution that arises from the
interparticle interactions. Equation \eqref{EQJ2inDDFT} is the natural
extension of the approximate one-body current \eqref{ddftCurrent} to
the two-body level.  External forces enter \eqref{EQJ2inDDFT} only
implicitly via the one-body density and current obtained by solving
\eqref{continuity1} with \eqref{ddftCurrent}. When combined with
\eqref{JtwoContinuity1}, equation \eqref{EQJ2inDDFT} provides a closed
equation of motion for the back van Hove current.

Taking the divergence with respect to $\rv_3$ in \eqref{EQJ2inDDFT},
using the two-body continuity equations \eqref{continuity1} and
  \eqref{JtwoContinuity2}, and integrating the entire history up to
$t_3$ yields an approximation to the front van Hove current
\begin{align}\label{TwoBodyCurrentEquation} 
\J^{\rm f,ad}_{\rm vH}(1,3)&= \J(1)\GvH(1,3) -D_0\rho(1) \nabla_1 
\Big(\GvH(1,3)
\notag\\&\hspace*{-0.85cm}
-\int d\rv_2 c(1,2_1) \rho(2_1) \left(\GvH(2_1,3)\!  
-\! \rho(3_{-\infty})\right)\Big),
\end{align} 
where $\rho(3_{-\infty})\equiv \rho(\rv_3,-\infty)$, a contribution
$\nabla_1 \rho(3_{-\infty})$ vanishes, and we have made the assumption
that two-body correlations factorize for widely separated time
arguments, i.e.  $\langle \hat{\rho}(\rv,t)\hat{\rho}(\rv',-\infty)
\rangle=\rho(\rv,t)\rho(\rv',-\infty)$, which holds in the absence of
an ideal glass transition \cite{reichman}.  Substitution of
\eqref{TwoBodyCurrentEquation} into the two-body continuity equation
\eqref{continuity2} yields a closed equation for the van Hove
function, which is local in time but nonlocal in space.  Equation
\eqref{TwoBodyCurrentEquation} is identical to the result for the
front van Hove current \cite{brader13noz}, derived by differentiating
\eqref{ddftCurrent} with respect to the one-body field \eqref{curlyV}.
This consistency demonstrates the flexibility of the method of
dynamical functional differentiation.

\subsection{Superadiabatic contributions}\label{beyondDDFT}
An exact theory for the two-body current should include a dependence
on the history of both the one- and two-body correlation functions.
Splitting the full two-body current into an adiabatic contribution
\eqref{EQJ2inDDFT} and a superadiabatic part, $\Jtwo=\Jtwo^{\rm
  ad}+\Jtwo^{\rm sup}$, the most general expression for the latter is
given by
\begin{align}\label{general_eom}
  &\Jtwo^{\rm sup}(1,3)=
  \rho(1){\sf M}(1,3)\rho(3)\\  & \quad
    +\rho(1) \int d2 \Big({\sf M}(1,2)\cdot\Jtwo(2,3)
    +{\bf m}(1,2)\JvHb(2,3)\Big),\notag
\end{align}
where we have introduced vectorial and tensorial direct time
correlation functions, denoted by ${\bf m}(1,2)$ and ${\sf M}(1,2)$,
respectively.  The two-body current thus consists of a direct
contribution, proportional to ${\sf M}(1,2)$, and a convolution
contribution of the respective direct time correlation function and a
probabilistic correlation function. The non-Markovian equation
\eqref{general_eom} is the most general form that involves only one-
and two-body functions which generate a tensor field from spacetime
convolutions of the van Hove current and the two-body current.  The
$2$-integral in \eqref{general_eom} runs over a spacetime slab from
the earlier time $t_3$ to later time $t_1$, consistent with causality.
Unlike the approximate expression \eqref{TwoBodyCurrentEquation}, the
exact NOZ equation \eqref{general_eom} is not closed and serves to
define the direct time correlation functions ${\bf m}(1,2)$ and ${\sf
  M}(1,2)$.

As an aside, we note the close structural similarity between
\eqref{general_eom} and the equilibrium OZ equation \cite{Hansen06}
\begin{align}
h(\rv_1,\rv_3)=c(\rv_1,\rv_3)\, + \int \!d\rv_2\,\,
h(\rv_1,\rv_2)\rho(\rv_2)c(\rv_2,\rv_3),
\label{ozInhomogeneous}
\end{align}
which defines the static direct correlation function,
$c(\rv_1,\rv_2)$, in terms of the one-body density and the two-point
correlation function $h(\rv_1,\rv_2)=\langle
\hat\rho(\rv_1)\hat\rho(\rv_2) \rangle/(\rho(\rv_1)\rho(\rv_2))-1$.

Building the divergence of the superadiabatic two-body current with
respect to $\rv_3$ and integrating in time $t_3$ yields
\begin{align}
  \label{TwoBodyCurrentEquationGeneral}
  &\JvH^{\rm f,sup}(1,3)=\JvH^{\rm f,sup}(1,3_{-\infty})
  -\rho(1)\!\int_{-\infty}^{\,t_3} \!\!\!dt'_3\, \nabla_3\cdot{\sf M}(1,3')\rho(3') 
  \notag\\  & \quad
    +\rho(1)\! \int \!d2\, \Big[{\sf M}(1,2)\cdot(\JvHf(2,3)-\J(2)\rho(3_{-\infty}))\notag\\
    & \quad
    \hspace*{1.4cm}+{\bf m}(1,2)\rho(2)(\GvH(2,3)-\rho(3_{-\infty}))\Big],
\end{align}
where $\JvHf=\JvHfddft+\JvH^{\rm f,sup}$.  Closure of the theory
requires \eqref{TwoBodyCurrentEquationGeneral} to be supplemented by
two independent equations that relate ${\bf m}(1,2)$ and ${\sf
  M}(1,2)$ to the van Hove function and its front current. One
possible approach is to postulate a dynamical closure relation, in
analogy to the procedure employed in equilibrium to obtain, e.g.\ the
Percus-Yevick or hyper-netted-chain approximations \cite{Hansen06}.
In Ref.~\cite{brader13noz} it was demonstrated that MCT can be viewed as a
closure of this type.  Alternatively, the power functional formalism
\cite{power} can be exploited to relate the time-direct correlation
functions to a generating functional, as outlined in the following.

\subsection{Connection to power functional theory}\label{TheConnection}
The power functional theory \cite{power} is a recent approach to
describing the dynamics of interacting Brownian systems by extending
classical density functional theory to nonequilibrium situations.  The
theory rests on a dynamic generating functional (the free power),
which is minimized with respect to either the current or the density,
yielding a pair of complementary Euler-Lagrange equations.  While the
formulation is exact, obtaining a closed expression for the current
requires knowledge of the excess (over ideal gas) power dissipation,
$P_t^{\rm exc}[\rho,\J]$, which is a functional of the history of
$\rho(1)$ and $\J(1)$ prior to time~$t$ and accounts for dissipation
that arises from interparticle interactions.

\subsubsection{Euler-Lagrange equation for the current}
Minimization of the dynamic free power functional with respect to the
current yields a general and exact equation of motion \cite{power}
\begin{align}
 \J(1) &=
 \J_{\rm DDFT}(1) - \frac{\rho(1)}{\gamma}
  \frac{\delta P_{t_1}^{\rm exc}[\rho,\J]}{\delta \J(1)},
  \label{EQcurrentPFT}
\end{align}
where $\J_{\rm DDFT}(1)$ is defined via \eqref{ddftCurrent} and arises
from differentiation of the ideal gas contribution to the power
dissipation, $P_t^{\rm id}[\rho,\J]=\int d\rv \gamma
\J(\rv,t)^2/(2\rho(\rv,t))$, where the total free power functional
consists of the sum $P_t[\rho,\J]= P_t^{\rm id}[\rho,\J]+ P_t^{\rm
  exc}[\rho,\J]$.
  
Differentiating \eqref{EQcurrentPFT} with respect to $\Xv(3)$,
  observing \eqref{EQJbyX}, and comparing the result to the general
form~\eqref{general_eom} yields the identification
of the direct time correlation functions with second functional
derivatives of the excess power dissipation via
\begin{align}
  {\bf m}(1,2) &= -\gamma^{-1}\frac{\delta}{\delta \rho(2)} 
  \frac{\delta P_{t_1}^{\rm exc}[\rho,\J]}{\delta \J(1)},
  \label{EQdefinitionDTCFvector}\\
  {\sf M}(1,2)^{\sf T} &= -\gamma^{-1}\frac{\delta}{\delta \J(2)} 
  \frac{\delta P_{t_1}^{\rm exc}[\rho,\J]}{\delta \J(1)},
  \label{EQdefinitionDTCFmatrix}
\end{align}
where the superscript $\sf T$ indicates the transpose.  The relations
\eqref{EQdefinitionDTCFvector} and \eqref{EQdefinitionDTCFmatrix} are
identical to those previously derived in Ref.~\cite{brader13noz} via the
less general method of differentiating with respect to the field
\eqref{curlyV}.  Equations \eqref{TwoBodyCurrentEquationGeneral} and
\eqref{EQcurrentPFT}--\eqref{EQdefinitionDTCFmatrix} show that only a
single mathematical object, the excess power dissipation functional,
need be approximated to generate a closed and consistent set of
equations for the dynamics of both the one- and two-body correlation
functions.  The DDFT \eqref{ddftCurrent} and the corresponding
adiabatic approximation \eqref{TwoBodyCurrentEquation} are obtained by
setting $P_t^{\rm exc}[\rho,\J]=0$. See Ref.~\cite{brader13noz} for
the discussion of a variety of systematic approximations beyond this
dynamical ideal gas approximation.

\subsubsection{Euler-Lagrange equation for the density}
The generality of the theory that we have developed enables us to
formulate a complete set of NOZ equations by observing that the power
functional framework contains, in addition to \eqref{EQcurrentPFT}, a
second Euler-Lagrange equation \cite{power}
\begin{align}\label{second_EL}
  -\frac{\gamma\J(1)^2}{\rho(1)^2} + 
  \frac{\delta P_{t_1}^{\rm exc}[\rho,\J]}{\delta \rho(1)}
  +\frac{\delta \dot F[\rho]}{\delta \rho(1)}=
  \alpha(1)-\dot V_{\rm ext}(1), 
\end{align}
which follows from minimization of the free power with respect to the
density distribution.  As already mentioned in
Sec.~\ref{DynamicalDerivatives}, the one-body function $\alpha(1)$
acts as a thermostat controlling the rate at which thermal energy is
either removed from or input to the system at a given spatial
location. The adiabatic approximation to \eqref{second_EL}, analogous
to the DDFT, is obtained by setting $P_t^{\rm exc}[\rho,\J]=0$.

The linearity of \eqref{second_EL} in $\dot V_{\rm ext}(\rv,t)$
suggests that the most natural way to generate an expression for the
van Hove current is to differentiate \eqref{second_EL} with respect to
$\dot V_{\rm ext}(\rv,t)$, which is the field conjugate to the
one-body density. Differentiating the adiabatic approximation to
\eqref{second_EL} with respect to $\dot V_{\rm ext}(3)$ thus yields an
approximate expression for the van Hove current
\begin{align}
\label{EQSecondNOZwithinDDFT}
&\vel(1)\cdot\J^{\rm f, ad}_{\rm vH}(1,3)= \vel(1)\cdot\J(1)\GvH(1,3)\\
&-D_0\rho(1) \frac{\partial}{\partial t_1}\Big(\GvH(1,3) 
 -\int \!d\rv_2 \, c(1,2_1) \rho(2_1) \GvH(2_1,3)\Big),\notag
\end{align}
which complements our previous result \eqref{TwoBodyCurrentEquation}
arising from the Euler-Lagrange equation for the current.  If we now
split, as we did previously, the van Hove current into adiabatic and
superadiabatic contributions, $\JvHf=\JvHfddft+\JvH^{\rm f,sup}$, the
superadiabatic contribution satisfies
\begin{align}
  \label{NEWsecondFullNOZ}
 &\vel(1)\cdot\JvH^{\rm f,sup}(1,3)=
  \J(1)\cdot{\bf m}(1,3)\rho(3)\\  & 
    +\rho(1)\!\!
    \int\!\! d2 \Big({\bf m}_b(1,2)\cdot\JvHf(2,3)
    +m_0(1,2) \rho(2)\GvH(2,3)\Big).\notag
\end{align}
Here the additional direct time correlation
functions are given as functional derivatives of the excess
dissipation functional
\begin{align}
  m_0(1,2) &= -\gamma^{-1}\frac{\delta}{\delta \rho(2)}
  \frac{P_{t_1}^{\rm exc}[\rho,\J]}{\delta \rho(1)},\label{second_mem1}\\
  {\bf m}_b(1,2) &= -\gamma^{-1}\frac{\delta}{\delta\J(2)}
  \frac{P_{t_1}^{\rm exc}[\rho,\J]}{\delta\rho(1)}.\label{second_mem2}
\end{align}
Knowledge of the excess dissipation functional is thus sufficient to
generate all direct time correlation functions.

The Euler-Lagrange equation \eqref{second_EL} can also be used
to generate a vectorial expression involving the two-body current.
Differentiating \eqref{second_EL} with respect to $\Xv(3)$ yields the
adiabatic contribution
\begin{align} 
&\vel(1)\cdot\Jtwo^{\rm ad}(1,3)= \vel(1)\cdot\J(1)\JvHb(1,3)\\
&-D_0\rho(1) \frac{\partial}{\partial t_3}\Big(\JvHb(1,3) 
 -\int d\rv_2 c(1,2_1) \rho(2_1) \JvHb(2_1,3)\Big),\notag
\end{align}
and the superadiabatic part 
\begin{align}\label{irreducible_second}
 &\nabla_3\vel(1):\Jtwo^{\rm sup}(1,3)=
  -\J(1)\cdot\frac{\partial}{\partial t_3}{\bf m}(1,3)\rho(3)\\  & 
    +\rho(1)\nabla_3\cdot\!\!
    \int\!\! d2 \Big({\bf m}_b(1,2)\cdot\Jtwo(2,3)\notag\\&\hspace{25mm}
    +m_0(1,2) \rho(2)\JvHb(2,3)\Big).\notag
\end{align}
We thus have a obtained complete set of NOZ equations for the two-time
correlation functions, based on both Euler-Lagrange equations.

\section{Concluding remarks}\label{conclusions}
To summarize, we have presented a method for calculating dynamical
functional derivatives and have applied this to derive general
expressions for the two-body correlation functions, most significantly
the tensorial two-body current. Our approach to calculating two-body
correlations is based on differentiation of the propagator for the
time evolution of the many-body distribution.  Importantly, the
dynamical operator which has to be considered when performing these
variational calculations is not the physical Smoluchowski operator
\eqref{EQomegaDefinition}, but rather a more general sourced time
evolution operator \eqref{EQomegaFree}, which is only {\it a
  posteriori} constrained by the many-body continuity equation
\eqref{smoluchowski}. The fact that the physical dynamics are enforced
after performing the variation is in keeping with the spirit of the
power functional theory \cite{power}, whereby derivatives with respect
to the current are performed at fixed density and vice versa.
 
When applied to the approximate DDFT one-body current
\eqref{ddftCurrent} the scheme yields an explicit, adiabatic
approximation to the two-body current \eqref{EQJ2inDDFT}. Time
integration of the divergence of the two-body current then yields an
expression for the front van Hove current and thus, via
\eqref{continuity2}, a closed expression for the van Hove function.
Going beyond the adiabatic approximation and incorporating the physics
of structural relaxation necessitates the introduction of time-direct
correlation functions.  The general result for the two-body current
consists of the sum of adiabatic \eqref{EQJ2inDDFT} and superadiabatic
\eqref{general_eom} contributions.  In order to formulate a closed
theory additional expressions for the two time-direct correlation
functions, ${\bf m}(1,2)$ and ${\sf M}(1,2)$, are required.

A possible way to apply the equations derived in this work is to view
the general NOZ theory as a basis for the construction of
nonequilibrium integral equation theories. However, a possibly
preferable approach is provided by the power functional formalism
\cite{power}, which enables the time-direct correlation functions to
be identified as functional derivatives of the excess power
dissipation functional. The challenge is to then find a suitable
approximation to this fundamental generating functional,
cf.\ Ref.~\cite{power}. As the same excess power dissipation generates
the dynamics of the one-body fields, the power functional approach can
be viewed as providing a unified (variational) framework for the
calculation of one- and two-body dynamical correlation functions, both
in and out of equilibrium.

In making a connection to the power functional theory~\cite{power} we
have used the pair of Euler-Langrange equations, which are obtained by
minimizing the free power either with respect to the one-body current
or the one-body density.  Although both Euler-Lagrange equations
provide expressions for the current and have an equivalent formal
status, they differ in interpretation.  The equation for the current,
which was addressed also in Ref.~\cite{brader13noz}, provides the most
straightforward extension of the DDFT result \eqref{ddftCurrent}, by
supplementing this with an additional term arising from
interaction-induced dissipation.  The Euler-Lagrange equation
\eqref{second_EL} for the density describes the local rate of energy
changes in the system.  The right hand side of \eqref{second_EL}
involves non-mechanical work arising from changes in the external
potential at a given spacetime point, as well as energy injection or
removal by a thermostat, as represented by the function $\alpha(1)$.
Given that we are able to distinguish between various forms of work,
it may prove fruitful to explore connections between the power
functional formalism and fluctuation theory
\cite{RegueraReiss,SeifertSpeck,SeifertReview}.

In Section \ref{ResponseFunctions} we interpreted the functional
derivatives \eqref{EQJbyX}-\eqref{EQrhoByRho} as dynamic nonlinear
response functions.  This led directly to equations
\eqref{responseJfromX}, \eqref{responseRhoFromVextdot},
\eqref{responseRhoFromX}, and \eqref{responseJfromVextdot}, which
demonstrate how the two-time correlators mediate external field
perturbations into resulting changes in the one-body fields.  These
Green-Kubo-type formulae for the current and density response follow
in a natural way from application of the dynamical functional
calculus.  However, the results presented here form only part of a
more general picture.  We expect to provide a unified variational
framework for the generation of formal Green-Kubo relations for
transport coefficients. Moreover, when applied to calculate the stress
response to mechanical deformation a formal constitutive relation can
be derived in a straightforward and physically intuitive fashion
\cite{brader10review}.  A full investigation of the connections
between the present approach and the established Green-Kubo formalism
\cite{kubo,kubobook,ForsterBook,ChaikinLubensky} is currently
underway.

We have focused here on the special case of Brownian many-body
dynamics. However, the functional differentiation of the dynamical
propagator to generate two-body correlation functions, is by no means
limited to this particular choice of microscopic dynamics, but rather
presents a general and powerful formalism. The application to
Newtonian dynamics, as described by the Liouville propagator
\cite{Hansen06}, should prove very instructive and may open the door
to new studies of dynamical processes in atomic liquids. Work along
these lines is also in progress.


\begin{thebibliography}{10}

\bibitem{evans79}
R. Evans, Adv. Phys. {\bf 28},  143  (1979).

\bibitem{vanHove54}
L. van Hove, Phys. Rev. {\bf 95}, 249 (1954).

\bibitem{Hansen06}
J.~P. Hansen and I.~R. McDonald, {\it Theory of Simple Liquids}, 3rd ed.
  (Academic Press, London, 2006).

\bibitem{brader13noz}
J.~M. Brader and M. Schmidt, J. Chem. Phys. {\bf 139}, 104108 (2013).

\bibitem{goetze_book}
W. G{\"o}tze, {\it Complex Dynamics of Glass-Forming Liquids: A Mode-Coupling
  Theory} (Oxford University Press, Oxford, 2009).
  
\bibitem{dtplhopkins10dtpl}
A.~J. Archer, P. Hopkins, and M. Schmidt, Phys. Rev. E {\bf 75},  040501(R)
  (2007);
P. Hopkins, A. Fortini, A.~J. Archer, and M. Schmidt, J. Chem. Phys. {\bf 133},
   224505  (2010).  
   
\bibitem{power}
M. Schmidt and J.~M. Brader, J. Chem. Phys. {\bf 138} 214101 (2013).   

\bibitem{dhont}
J. K. G.~Dhont, {\it An Introduction to the Dynamics of Colloids}
(Elsevier, Amsterdam, 1996).

\bibitem{vankampen}
N.G.~van Kampen, 
{\it Stochastic Processes in Physics and Chemistry}
(North-Holland, Amsterdam, New York, Oxford 1981).

\bibitem{braderMCT2012}
J.~M. Brader, M.~E. Cates, and M. Fuchs, Phys. Rev. E {\bf 86},  021403
  (2012).
  
\bibitem{riskin}
H. Risken, {\it The Fokker-Planck Equation} 
(Springer, 1996).

\bibitem{larson}
R. G.~Larson, {\it The Structure and Rheology of Complex fluids} 
(Oxford, 1998).

\bibitem{marinibettolomarconi99}
U. {Marini Bettolo Marconi} and P. Tarazona, J. Chem. Phys. {\bf 110},  8032
  (1999).

\bibitem{archer04ddft}
A. J.  Archer and R. Evans, J. Chem. Phys, {\bf  121}, 4246 (2004).

\bibitem{zwanzig_book}
R. Zwanzig, {\it Nonequilibrium Statistical Mechanics} (Oxford University
  Press, Oxford, 2001).

\bibitem{reichman}
D. R.~Reichman and P.~Charbonneau, J. Stat. Mech. P05013 (2005) 

\bibitem{RegueraReiss}
D.~Reguera and H.~Reiss, J. Chem. Phys. {\bf 120}, 2558 (2004).

\bibitem{SeifertSpeck}
T.~Leonard, B.~Lander, U.~Seifert and T.~Speck, 
J. Chem. Phys. {\bf 139}, 204109 (2013).

\bibitem{SeifertReview} 
U.~Seifert, Rep. Prog. Phys. {\bf 75}, 126001 (2012).

\bibitem{brader10review}
J.~M. Brader, J. Phys.: Condens. Matter {\bf 22}, 363101 (2010).

\bibitem{kubo}
M. S. Green, J. Chem. Phys {\bf 22}, 398 (1954);
R. Kubo, J. Phys. Soc. Jpn. {\bf 12} 570 (1957). 

\bibitem{kubobook}
R. Kubo, M. Toda, and N. Hashitsumi,
{\it Stastical Physics II: Nonequilibrium Statistical Mechanics}
(Springer, Berlin, 1991).

\bibitem{ForsterBook} 
D. Forster {\it Hydrodynamic Fluctuations, Broken Symmetry and Correlation Functions} 
(Benjamin, Reading, MA, 1975).

\bibitem{ChaikinLubensky} 
P.M. Chaikin and T.C. Lubensky, 
{\it Principles of Condensed Matter Physics} 
(Cambridge University Press, Cambridge, 1995).  

\end{thebibliography}
\end{document}